\documentclass[10pt,letterpaper,twocolumn]{article}
\usepackage{ol2}
\usepackage{varioref}
\usepackage{graphicx}
\usepackage[T1]{fontenc}
\usepackage{subfigure}
\usepackage{cite}
\usepackage{amsmath}
\usepackage{amsfonts}
\usepackage{stmaryrd}
\usepackage{psfrag}
\usepackage{tabularx}\newcolumntype{Y}{>{\centering\arraybackslash}X}% %% pour centrer les textes dans tabularx
\graphicspath{{/home/image/eps/pmma/}}
\DeclareGraphicsExtensions{.eps,.EPS,.ps,.PS,.eps.gz}
\begin{document}

\twocolumn[
\title{Ultracompact fiber integrated X-ray dosimeter based on scintillators coupled to a nano-optical antenna.}

\author{Zhihua Xie, Hichem Maradj, Miguel Suarez, Lydie Viau, Virgine Moutarlier, Carole Fauquet, Didier Tonneau and Thierry Grosjean}

\address{Optics Department - FEMTO-ST Institute UMR 6174 - Univ. Bourgogne Franche-Comte - CNRS - Besancon, France}

\address{Department of Nano-object Sciences and Technologies - CINaM Institute UMR 7325 - Univ. Aix Marseille - CNRS - Marseille, France}

\address{Department of Structured Materials and Surfaces - UTINAM Institute UMR 6213 - Univ. Bourgogne Franche-Comte - CNRS - Besancon}

\email{thierry.grosjean@univ-fcomte.fr}

\begin{abstract}
High energy radiations are nowadays cornerstone in scientific, medical and industrial imaging and cancer therapy. However, their detection and dosimetry are restricted to systems of modest compactness that are not compatible with endoscopy. Here, we report the first experimental imaging and dosimetry of X-rays with an ultracompact sensor integrated at the end of a single mode optical fiber (125 $\mu$m full diameter). We realize such a dosimeter by coupling luminescent scintillators to a nano-optical antenna engineered at the end of the fiber and by detecting in-fiber outcoupled luminescence with a photon counter. Direct profile and real-time dosimetry of a X-ray focused  beam reveals a spatial resolution of a few microns together with a sensitivity better than 10$^3$ X-photons$/$s$/\mu$m$^2$ at energies of 8-10 keV.  Because the nano-optical platform is auto-aligned with respect to the fiber, the resulting dosimeter is plug-and-play, flexible, and it is suitable for ultralow footprint endoscopic investigations at higher X-ray energies. Our nano-optical approach thus offers the  possibility of X-ray profiling and dosimetry in ultra-confined environments, opening up new avenues in the fields of X-ray imaging, real-time control of cancer radiotherapies and Curie therapies, medical and industrial endoscopy, etc. With this study, nano-optical antennas make a first key contribution to the development of X-ray sensing protocols and architectures. 
\end{abstract}

\maketitle ]

\section{Introduction}

Nano-optical antennas (NOAs) are becoming pivotal tools in modern photonics owing to their unprecedented ability to control light-matter interaction in ultracompact architectures \cite{novotny:natphot11}. Various NOAs concepts, inherited from low frequency antenna theory have for instance demonstrated unprecedented ability to tailor emission rate and directionality from luminescent elemental sources such as fluorescent molecules and quantum dots \cite{claudon:natphot10,curto:science10,farahani:prl05,muskens:nl07,belacel:nl13,devilez:acsnano10,lu:acsnano12}. We address here the use of a NOA concept for controlling the X-ray excited luminescence (XEL) from scintillators, thus providing a nano-optically driven approach in the development of novel architectures for X-ray imaging and real-time dosimetry.  

The development of miniaturized X-ray sensors and dosimeters is hindered by the difficulty to achieve efficient "`X-photon"'-to-electron conversion in electronic devices, which imposes large detection volumes. For instance,  ionization chambers used for measuring X-ray dose in radiotherapy show detection volume larger than cm$^3$\cite{hine:book}. Indirect detection, which combines luminescent materials to visible optical detectors, has demonstrated performances in terms of image contrast and signal dynamics, and is now widely exploited in a large panel of scientific, medical and industrial domains. In such a technique, semiconductor materials used under various forms, such as crystals or powders, (called scintillators or phosphors) convert high energy impinging radiations into light that is detected with silicon-based photodiodes and cameras or with photosensitive films. The resulting optical devices are often of modest compactness, which may represent limitations from practical point-of-view. However, the technique offers great promises in low X-ray flux detection since one can detect optically luminescence signal down to a single photon. 

Accessing direct in-fiber XEL detection offers the prospect of wide range of X-ray sensors and dosimeters free from bulky optics. The integration of X-ray detection functionality at the end of an optical fiber is highly desirable as it would lead to ultra-compact, plug-and-play and flexible architectures allowing for a completely new versatility in X-ray imaging and dosimetry. In the medical domain, real-time X-ray dosimetry right at a targeted tumor would be possible with endoscopic techniques, thus enabling enhanced accuracy and control in radiotherapy and remarkably improved performances in cancer treatment. 

Reaching efficient optical coupling between scintillators and conventional step index optical fibers remains a real challenge owing to the strong mismatch between the almost omnidirectional dipolar emission of luminescent particles and the noticeably low numerical aperture and weak guiding properties of the fibers. The approach followed so far consists of compensating this low coupling efficiency by considering large scintillating volume coupled to large core multimode fibers \cite{archambault:medphys07,beddar:pmb92,beddar:rm06,moon:ari12,letourneau:mp99,lee:tns08,lee:kem06}. These techniques lead to modest resolution on the mm to cm range and a compactness weakly compatible with endoscopy. Recently, sub-millimeter architecture has been demonstrated by covering the cleaved end facet of a 600 $\mu$m core diameter multimode fiber with a 11 $\mu$m thick layer of scintillators \cite{belley:mp15}. Unfortunately, the high aspect ratio of the sensor leads to important spatial resolution anisotropy in X-ray beam profiling and dosimetry. Scaling down dosimeter architecture with a direct scintillator-to-fiber coupling, for instance by improving optical detection with photon counters, remains questionable and has never been addressed yet. We recently theoretically proposed an alternative approach relying on the engineering of an ultracompact interface between a luminescent source and a fiber, aimed at optimizing the in-fiber photon outcoupling \cite{grosjean:ox13}. This interface is the result of the concept of horn antenna to optical frequencies, capable of collecting and transferring up to 70\% of the luminescence photons to the fiber guided mode. 

In this letter, we use this NOA approach to show experimentally direct plug-and-play imaging and dosimetry of a X-ray focused beam with a spatial resolution of a few microns.  By optimizing the coupling channel between a tiny scintillation cluster and an optical fiber, we achieve highly miniaturized and spatially isotropic X-ray detection systems: X-ray detection volumes of a few tens of $\mu$m$^3$ can be obtained at the end of a 125 $\mu$m or even 80 $\mu$m diameter single mode fiber, thus noticeably increasing both system overall compactness and measurement accuracy. This opens up perspectives of ultracompact, flexible, plug-and-play and high resolution real-time dosimeter for monitoring, mapping and controlling the dose of high energy radiations in a wide panel of applications covering scientific, medical and industrial domains.

\section{Principle and fabrication}

From antenna theory \cite{balanis:book}, two ways are possible for impedance matching the emission from a point-like electromagnetic source, such as the end of a coaxial cable, with vacuum. The first way relies on the generation of localized resonances over subwavelength assemblies of metallic elements of specific shapes and sizes. Such a concept has recently been transposed to optics and led to well-known concepts of resonant NOAs \cite{novotny:natphot11}. The second way is based on the efficient coupling of the point-like source to a flaring waveguide, leading to larger antenna structures such as the horn antenna which has also been recently successfully extended to optical frequencies but remains relatively unexplored \cite{ramaccia:ol11,yang:ol14,grosjean:ox13}. A horn antenna usually directs emission from a dipolar source in free space, by connecting it to a coax-to-waveguide adapter, resulting in a simple architecture where the point-like emitter is placed in between a reflector and the flaring waveguide. This basic concept, which has been already transposed to optics \cite{grosjean:ox13}, is used in the present work to optically harness a scintillator cluster  to a single mode fiber. 

%%%%%%%%%%%%%%%%%%%%%%%%%%%%%%%%%%%%%%%%%%%%%%%%%%%%%%
\begin{figure}[htbp]
\centering
\fbox{\includegraphics [width=\linewidth]{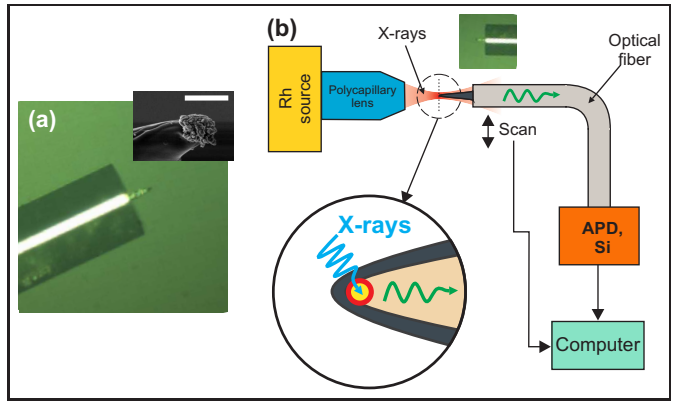}}
\caption{(a) Optical microscope image of a fiber-integrated NOA based X-ray dosimeter. Inset: SEM micrograph of a scintillation cluster at the apex the micro-tip before aluminum coating (scalebar: 4 $\mu$). (b) Scheme of the experimental set-up.}\label{fig:fab_scheme}
\end{figure}
%%%%%%%%%%%%%%%%%%%%%%%%%%%%%%%%%%%%%%%%%%%%%%%%%%%%%%

In order to produce the scintillator-coupled horn NOA fiber system, polymer micro-tips are first grown by photopolymerization at the cleaved end facet of a single mode fiber (SMF-28) \cite{bachelot:ao01}. Tips are about 30 $\mu$m long and have a radius of curvature of about 1 $\mu$m at their apex. Next, a scintillator cluster (barium platino-cyanide) is attached at the micro-tip end under microscope observation, following the approach proposed in Ref. \cite{aigouy:ao04}. Isolated micron-size clusters (between 1 and 5 $\mu$m) are deposited onto a flat surface and the micro-tip is approached to a single cluster until contact. Owing to adhesion forces between tip and cluster, the luminescent material is kept attached to the micro-tip when it is removed from the surface.  Finally, the resulting fiber-integrated structure is metal coated with a few nanometer thick titanium adhesion layer followed by an aluminum layer that is 100-150 nm thick. Aluminum is chosen for its high reflectivity at visible wavelengths and high transparency to X-rays. Figure \ref{fig:fab_scheme}(a) displays optical and SEM micrographs of a resulting fibered X-ray dosimetry platform. Because the SEM irradiation tends to burst the polymer microtip (see inset of Fig.  \ref{fig:fab_scheme}(a)), which irremediably results in a drop of luminescence collection efficiency, our scintillator grafted fiber micro-tips used in the following experiments have not been characterized by SEM, leading to uncertainty in the estimation of the cluster size and thus of the spatial resolution ability of the probe. Note that we consider here luminescent clusters larger than the wavelength embedded within the NOA, instead of single dipolar emitter as mentioned in Ref. \cite{grosjean:ox13}. The distribution of crystalline defect centers within the cluster (promoters of X-ray-to-light conversion in our case \cite{lecoq:book}) can be considered as an ensemble of optical dipolar sources coupled to the antenna, and the horn NOA concepts remains valid in that situation and demonstrates high performances in X-ray detection.

\section{Results and discussion}

Our concept of ultra-compact fiber-integrated X-ray dosimeter is first demonstrated with the fully calibrated X-ray focusing bench shown in Fig. \ref{fig:fab_scheme}\cite{dehlinger:nrl13}. Polychromatic X-ray radiation from a low power Rh target lab source (800 $\mu$A at 35 kV) is focused with a polycapillary lens into a 25 $\mu$m wide spot (FWHM, manufacturer's data). The X-ray fiber dosimeter is positioned right at the focus with a 3D manual translation stage. The final centering process together with beam scanning is realized with a precision motorized system. During scanning, X-rays cross the transparent tiny dosimeter with minimum perturbations, the thin aluminum layer does not prevent the primary X-ray beam from directly exciting the NOA-embedded scintillators.  Owing to high reflectivity of aluminum at visible wavelengths, a large portion of the XEL from the scintillators is collected by the NOA and launched towards the fiber core with high directionality. The unique impedance matching property of flaring waveguides, especially of our quasi-adiabatic tapered micro-tips initially produced by the fiber mode itself (during photo-polymerization process), ensures optimum photon outcoupling into the fiber mode. In-fiber XEL detection is realized with a room temperature photon counter (LynXea) from Aurea Technology.

%%%%%%%%%%%%%%%%%%%%%%%%%%%%%%%%%%%%%%%%%%%%%%%%%%%%%%
\begin{figure}[htbp]
\centering
\fbox{\includegraphics [width=0.99\columnwidth]{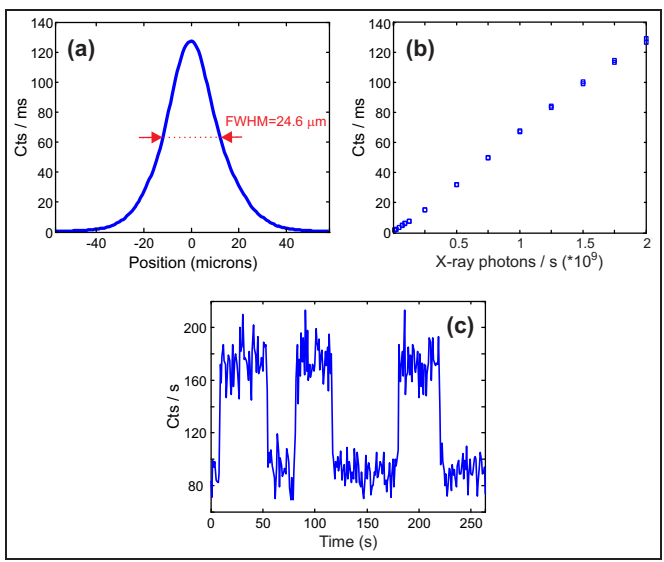}}
\caption{(a) Experimental plot of a X-ray focused beam (>8keV) at focus. (b) Calibration curve of the dosimeter (detected luminescence vs incident X-ray flux). Minimum detectable X-ray power density is here estimated to the order of 10$^3$ photons/s/$\mu$m$^2$. (c) Time trace of the luminescence intensity with the same ultra-compact fiber dosimeter positioned within collimated X-ray radiations emanating from a Cu target source of a commercial diffractometer (Bruker).}\label{fig:results}
\end{figure}
%%%%%%%%%%%%%%%%%%%%%%%%%%%%%%%%%%%%%%%%%%%%%%%%%%%%%%

Fig. \ref{fig:results}(a) shows X-ray beam profile at focal plane, revealing a full width at half maximum (FWHM) of 24.6 $\mu$m which is in good agreement with the manufacturer's spot size value. Note that, because both the beam focus and cluster size are much larger than the incident X-ray wavelengths, one can retrieve the real beam profile with direct deconvolution techniques between the experimental plot and the point-spread function of the fiber probe. Point spread function could be measured by probing the focal plane of a X-ray Fresnel lens (which show focal spot of the order of a few nanometers) \cite{chao:ox09} or could be deduced from the measurement of an edge response. Such an image processing is however beyond the scope of this paper, and will be studied later. According to the basic rules related to the convolution of gaussian functions, the experimentally measured FWHM very close to the manufacturer's reference value of 25 $\mu$m confirms that the luminescent source is on the micron scale, which is consistent with the cluster sizes measured by SEM during the preliminary tests of scintillator-to-tip grafting process (cf. Inset of Fig. \ref{fig:fab_scheme}(a)). Little drifts of the probe during image acquisition and possible rounding of the manufacturer's spot size value may also explain a part of the discrepancy observed here. 

The calibration curve of our X-ray dosimeter, shown in Fig. \ref{fig:results}(b), reveals a linear response of the system with respect to the incident X-ray power. For a maximum X-ray power of 2 10$^9$ photons/s, detected luminescence signal reaches 130 kcts/s, which dramatically exceeds the noise level of the photodetector.  This curve is used to evaluate the minimum X-ray power density detectable with our approach to a value smaller than 10$^3$ photons/s/$\mu$m$^2$. Our dosimeter is therefore sensitive enough to detect and probe X-rays from diffractometers. To validate this point, we performed direct X-ray dose measurement of low power Cu-target source used in a commercial  diffractometer from Bruker company (D8 Advance). Time trace of luminescence intensity with the source successively on and off is reported in Fig. \ref{fig:results}(c). We see that the detected signal is well beyond noise level, which validates our approach in low power X-ray radiation sensing. From Fig. \ref{fig:results}(b), the flux density of the diffractometer source is estimated at 3.7 10$^3$ photons/s/$\mu$m$^2$. Note that the correlated detection opportunity provided by photon counters \cite{sohier:phd11} offers the perspective of enhancing detection sensitivity while preserving or even increasing its resolution ability. Sensitivities of 300 photons/s/$\mu$m$^2$ have also been observed with the attachment of an ensemble of clusters onto a micro-tip, at the expense of a lower resolution ability (about 30 $\mu$m). 

X-ray irradiators available at clinical radiotherapy departments usually deliver dose rates of the order of a few centigrays per second (from 5.5 cGy/s to 40 cGy/s) with polychromatic X-ray radiations showing peak energy of a few hundreds of keV. In-fiber XEL detection has been already demonstrated under X-ray dose rates of the order 1.9 cGy/s with a 10 $\mu$m thick scintillator pellet covering the cleaved end facet of a 600 $\mu$m core fiber \cite{belley:mp15}. Optical powers of a few picowatts, i.e. about 10$^7$ photons/s, were then detected at the fiber output with a power-meter. Taking into account the dosimeter size reduction, X-ray absorption decrease of the scintillators (due to higher X-ray energies) and dramatic enhancement of XEL fiber outcoupling with our nano-optically driven technological approach, detected signals of the order of $10^3$ to $10^4$ visible photons per second are expected for a spatial resolution of 10-20 $\mu$m. This represents an important signal-to-noise ratio with photon counter monitoring. Our ultra-compact fiber sensor could thus be considered as an ultra-low footprint dosimeter for controlling in real-time radiotherapy processes. 

%%%%%%%%%%%%%%%%%%%%%%%%%%%%%%%%%%%%%%%%%%%%%%%%%%%%%%
\begin{figure}[htbp]
\centering
\fbox{\includegraphics [width=0.7\columnwidth]{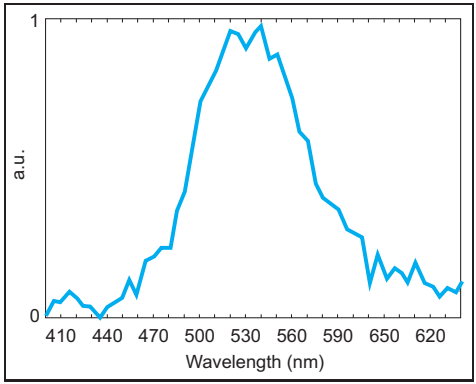}}
\caption{Spectrum of the detected optical signal when the fiber probe is centered with respect to the focus. }\label{fig:spectre}
\end{figure}
%%%%%%%%%%%%%%%%%%%%%%%%%%%%%%%%%%%%%%%%%%%%%%%%%%%%%%

We used for these experiments barium platino-cyanide as X-ray-to-light converter, which is known to produce XEL emission at a wavelength of 520 nm due to scintillation process promoted by its crystalline defect centers \cite{yen:book}. An abundant literature is however available on the fabrication and use of luminescent materials and the desired scintillator can be chosen in a large toolkit of commercialized products, depending on the targeted application \cite{nikl:mst06,yen:handbook,eijk:pmb02}. To ensure that the luminescence is induced by the scintillation cluster positioned at the end of the fiber micro-tip, we measured in-fiber collected luminescence spectrum from the scintillator-grafted NOA by connecting the fiber output facet to a spectrometer (Princeton SP2300) (see Fig. \ref{fig:spectre}). We see that the detected signal is spectrally centered to a wavelength of 530 nm, which is consistent with barium platino-cyanide emission spectrum. The little redshift of 10 nm of the emission peak with respect to the predicted spectrum may be due to the influence of the structured environment onto the cluster's emission properties. Moreover, we probed the X-ray focused beam with a fiber NOA free from scintillators. In that case, a detection signal of a few hundreds of luminescence photons per seconds was measured, which is much lower than the signal level obtained in the presence of the scintillation cluster (up to 130 kcts/s, see Fig. \ref{fig:results}(a,b)). These measurements show that the detected XEL is undoubtedly due to the scintillators embedded within the NOA.  

\section{Conclusion}

We proposed a nano-optically driven technological approach for the detection and real-time dosimetry of high energy radiations in ultracompact and flexible architectures. By coupling scintillators to a horn NOA and by exploiting the record emission directionality and impedance matching of this metallo-dielectric structure,  we successfully developed a micron-size X-ray sensor at the end of a 125 $\mu$m diameter single mode optical fiber in architecture of unprecedented compactness and flexibility. By leveraging the versatility and ubiquity of fiber-optics technology, this may constitute a key step towards the widespread use of ultracompact X-ray detector in a wide panel of scientific, medical and industrial domains.  For example, high energy imaging and real-time dosimetry (X-rays, gamma rays, charged particles, etc) for radiotherapy and Curie therapy would become possible with completely new versatility, accuracy and control possibilities, and with ultra-low footprint detection devices. More generally, NOAs makes with our approach a first key contribution to the development of X-ray sensing protocols and architectures. In that context, nano-optics would open the way towards radically new concepts allowing for smaller and faster X-ray detectors for scientific, medical and industrial metrology and characterization. While the capability of plasmonic NOAs to dramatically increase fluorescence decay rate has been intensively studied \cite{farahani:prl05,muskens:nl07,belacel:nl13}, their ability to control light-matter interaction generated under X-ray exposure has not been reported so far. NOAs thus also represent very promising  perspectives in the generation of faster X-ray detectors by enhancing radiative decay rate of scintillation processes.\\

This work is funded by the Labex ACTION (ANR-11-LABX-0001-01). The authors thank Lovalite company for technical support and are indebted to Claudine Filiatre for helpful discussions.

%\bibliography{base_biblio}
%\bibliographystyle{unsrt}

\end{document}